\documentstyle[aps,psfig,twocolumn]{revtex}

\newcommand{\ra}{\rightarrow}

\newcommand{\be}{\begin{equation}}
\newcommand{\ee}{\end{equation}}

\newcommand{\bea}{\begin{eqnarray}}
\newcommand{\eea}{\end{eqnarray}}
\newcommand{\bef}{\begin{figure}}
\newcommand{\eef}{\end{figure}}

\newcommand{\lgl}{\langle}
\newcommand{\rgl}{\rangle}

\tightenlines
\begin{document}
\draft
\title{Photons from Pb-Pb Collisions at CERN SPS}
\vskip 0.2 in
\author{Jan-e Alam$^{1,}$\cite{jane},
Sourav Sarkar$^2$, T. Hatsuda$^1$, Tapan K. Nayak$^2$
and Bikash Sinha$^{2,3}$}
\address{$^1$~Physics Department, University of Tokyo, Tokyo 113-0033, Japan}

\address{$^2$~Variable Energy Cyclotron Centre,
     1/AF Bidhan Nagar, Calcutta 700 064
     India}

\address{$^3$~ Saha Institute of Nuclear Physics,
           1/AF Bidhan Nagar, Calcutta 700 064
           India}

\maketitle
\begin{abstract}
High energy photon emission rate  
from matter created in Pb + Pb collisions at
CERN SPS energies is evaluated. 
The evolution of matter from the initial state up to freeze-out
has been treated within the framework of (3+1) dimensional
hydrodynamic expansion. We observe that the photon spectra
measured by the WA98 experiment are well reproduced
with hard QCD photons and photons from a thermal source 
with initial temperature~$\sim 200$ MeV. 
The effects of the spectral changes of hadrons 
with temperature on the photon emission 
rate and on the equation of state are studied.
Photon yield for Au + Au collisions at RHIC 
energies is also estimated.
\end{abstract}

\narrowtext

Ultra-Relativistic collisions of heavy nuclei have brought us
within reach of creating and studying various aspects 
of quark-gluon plasma (QGP), which
so far was believed to exist 
in the microsecond old universe or possibly 
in the cores of neutron or quark stars.
We are at a very interesting
situation in this area of research where the Super Proton
Synchrotron (SPS) era has drawn to a close and the first results from
the Relativistic Heavy Ion Collider (RHIC) have started to appear. 
Already from the results of the Pb run at the SPS quite a few of the 
signatures of QGP, {\it e.g.}, $J/\Psi$ suppression, strangeness enhancement
etc., are reported to have ``seen'' unmistakable hints 
of the existence of QGP~\cite{qm99}. 
Electromagnetic probes, {\it viz.}, photons
and dileptons have long been recognized as the most direct probes
of the collision~\cite{emprobe}. 
Owing to the nature of their interaction they undergo minimal scatterings
and are by far the best markers of the entire space-time evolution
of the collision.  

The single photon data, obtained from Pb-Pb collisions at CERN SPS
reported by the WA98 Collaboration~\cite{wa98}
have been the focus of considerable interest in recent 
times~\cite{prc1,cywong,kg,dyp,dks}. 
While in Ref.~\cite{prc1}, the contribution from hard photons
and the effects of transverse expansion on the thermal
photons were neglected, in \cite{cywong} the thermal
contribution was not taken into account.  Again, in
~\cite{kg,dyp} the
thermal shift on hadronic masses was neglected, whereas,
in~\cite{dks} both the 
the intrinsic $k_T$ distribution of partons and 
thermal shift of hadronic masses were ignored. 
In this letter we present the results of an analysis of
the photon spectra in a realistic framework 
using a  reasonable set of parameters and consistently taking 
into account all the effects mentioned above. 
We emphasize the effects of in-medium modifications of hadrons
on the photon spectra 
considering the fact that as yet it has not been possible
to explain the observed low-mass enhancement of dileptons measured
in the Pb+Au as well as S+Au  collisions at the CERN SPS in 
a scenario which does not incorporate in-medium effects 
on the vector meson mass (see ~\cite{rapp1} for a review).  

Let us first identify the
possible sources of ``excess'' photons above those coming
from the decays of pseudoscalar $\pi^0$ and $\eta$ mesons, as provided
by the data. Firstly, one has the prompt photons coming
from the hard collisions of initial state partons in the colliding
nuclei. These populate the high transverse momentum region 
and can be estimated by perturbative QCD.
The thermal contribution depends on the space-time evolution scenario
that one considers.
In the event of a deconfinement phase transition, one first has
a thermalized QGP which expands and cools, reverts back to hadronic matter,
again expands and cools and eventually freezes out into hadrons most of which 
are pions. Photon emission in the QGP occurs mainly due to QCD
annihilation and Compton processes between quarks and gluons.
In order to estimate the emission from the hadronic matter we
will consider a gas of light mesons {\it viz.} $\pi$, $\rho$, $\omega$
and $\eta$. 

It has been emphasized that the properties  of vector mesons 
may change appreciably because
of interactions among the hadrons at high temperatures and/or densities
(see the reviews~\cite{rapp1,thpr,brpr,rdp,annals}).
This modifies the rate of photon emission as
well as the equation of state (EOS) of the evolving matter. 
 Among various models for vector mesons available 
 in the literature~\cite{annals}, we examine two possibilities 
for the hadronic phase
  in this paper: (i) no medium modifications of hadrons, and 
   (ii) the scenario of the universal scaling 
   hypothesis of the vector meson masses~\cite{brpr}.
  In principle, we can think of a third scenario (iii)
   the large collisional broadening of the vector
    mesons~\cite{rapp1}.  Both (ii) and (iii) can reproduce the 
      enhancement of the low-mass dileptons measured by
       CERES Collaboration at CERN SPS, but the scenario (iii) has
been found to have a negligible effect 
on the emission rate of photons~\cite{annals}.
The effect of temperature dependent 
mass as described in case (ii) has also been incorporated
in the EOS of the hadronic matter 
undergoing a (3+1) dimensional expansion.  
We will see that the resulting photon spectra reproduce the 
experimental data quite well. 

There is still substantial debate on the order of the
phase transition as well as the value of the
critical temperature ($T_c$).
To address this aspect we will
also consider a scenario where
the system begins to evolve from a high temperature phase
where all the hadronic masses approach zero 
(pion mass is fixed at its vacuum value).
As the system expands and cools, the hadrons acquire
masses (as in case (ii) above) till freeze out.
Incorporation of medium modified masses and the EOS in this
case also provides a reasonable explanation of the data.

We start with the direct QCD photons originating from hard scattering 
of partons embedded in the nucleons of the 
colliding nuclei in the very early stages of the collision. 
We will see later that 
they make a significant contribution to the total photon yield. 
These are estimated using perturbative QCD as
\be
E\frac{dN}{d^3p}=T_{AA}(b=0)\,\,E\frac{d\sigma_{pp}}{d^3p}
\ee
where $T_{AA}(b)$ is the nuclear thickness at impact parameter $b$. Its value 
at $b=0$ is taken as 220/fm$^2$~\cite{wa98}. 
$\sigma_{pp}$ includes the $pp$ cross-section for
Compton and annihilation processes among the partons. At SPS energies one
should include the effects of intrinsic $k_T$ distribution
of partons~\cite{owens} to account for the fact that 
the colliding partons might have some initial transverse momenta with 
respect to the incoming hadrons (analogous to the Fermi motion of nucleons
in the nucleus). This leads to substantial enhancement 
in the photon spectra~\cite{cywong}. 
In practice such an effect is implemented by multiplying
each of the parton distribution functions appearing in the right hand side 
of the above equation by a Gaussian function of the type $f(k_{T})
=\exp[-k_T^2/\lgl k_T^2\rgl]/\pi\lgl k_T^2\rgl$ and integrating over
$d^2k_T$.
We use CTEQ5M partons~\cite{cteq} 
and $\lgl k_T^2\rgl=0.9$ GeV$^2$~\cite{cywong} for
evaluating the hard QCD photons ($\sqrt{<k_T^2>}=0.8$ GeV is taken in
\cite{kg}). 
The photons from hard QCD processes
have been used to normalize the p-p data. 
The higher order effects has been
taken into account through a K-factor $\sim 2$. 
An enhanced production in 
A-A collisions compared to p-p will presumably mark the presence of a 
thermal source~\cite{wa98}.

The lowest order processes contributing to hard thermal 
photon emission from quark
gluon plasma are the QCD Compton and annihilation 
processes~\cite{kapusta}. It has been
shown recently~\cite{aurenche} that the two-loop contribution leading to
bremsstrahlung and $q\bar q$ annihilation with scattering is of
the same order as the lowest order processes.. The total rate of emission
per unit four-volume at temperature $T$ is given by
\bea
E\frac{dR}{d^3p}&=&\frac{5}{9}\,\frac{\alpha\alpha_s}{2\pi^2}\,\,
\exp(-E/T)
\,\left[\ln\left(\frac{2.912\,E}{g^2\,T}\right)\right.
\nonumber\\
&&\left.+16\frac{(J_T-J_L)}{\pi^3}
\{\ln2+\frac{E}{3T}\}\right]
\label{edr}
\eea
where $J_T\simeq4.45$ and $J_L\simeq-4.26$. 
The QCD coupling, `$g$' is given by,
\be
g^2/4\pi\equiv\alpha_s=\frac{6\pi}{29\ln(8T/T_c)}
\label{strong}
\ee
for two quark flavours~\cite{fk}.
For photon energies in the range $1\leq$ $E$ (GeV) $\leq 5$, the static
emission rate given by Eq.~(\ref{edr}) 
at $T\sim 200$ MeV is about an order 
of magnitude larger than
the rate due to Compton and annihilation processes~\cite{kapusta}. We will see below that
within the present framework the space time integrated photon yield
from quark matter is less than that from hadronic
matter due to the smaller life time of the QGP phase as a
result of a moderate value of the initial temperature considered. 
Therefore,
the overall thermal photon yield remains largely unaffected by
the two loop contribution. 

To estimate the photon yield from the hadronic matter (HM)
(see first of~\cite{kapusta}), 
we have considered the reactions,
$\pi\,\rho\,\ra\, \pi\,\gamma$, 
$\pi\,\pi\,\ra\, \rho\,\gamma$, $\pi\,\pi\,\ra\, \eta\,\gamma$, 
$\pi\,\eta\,\ra\, \pi\,\gamma$ and the decays $\rho\,\ra\,\pi\,\pi\,\gamma$
and $\omega\,\ra\,\pi\,\gamma$.
The invariant amplitudes for all these 
processes are given in Refs.~\cite{npas}.
In the present work we have  also considered photon
production due to the process $\pi\,\rho\,\ra\,a_1\,\ra\,\pi\,\gamma
~\cite{annals}$. 

In Ref.~\cite{annals} we have studied the effects of spectral changes of hadrons
 on the electromagnetic probes in detail.
 It was observed that the gauged linear and non-linear  
sigma models and the model with hidden local symmetry do
 not show any appreciable effect on photon emissions.
 In the Walecka model, the universal scaling
 hypothesis for the vector meson masses as well as
   the large collisional broadening of vector mesons produce
  a large enhancement in low mass dileptons.
 However, the photon emission rate
  does not suffer substantial medium effect in the latter case, since
  the spectral function is smeared out.
  Nevertheless, the scaling hypothesis with particular exponent
   $\lambda=1/2$ (called the Nambu scaling in~\cite{brpr}) has been seen to
   enhance photon emission among the others.
   
 To consider the effect of the spectral modifications of hadrons
  we adopt
   two extreme cases: (i) no medium modifications of hadrons, and 
    (ii) the scaling hypothesis with  $\lambda=1/2$. 
In case (ii), the  parametrization of in-medium 
quantities (denoted by $*$) at finite $T$ is
\be
{m_{V}^* \over m_{V}}  = 
{f_{V}^* \over f_{V}} = 
{\omega_{0}^* \over \omega_{0}}  =
 \left( 1 - {T^2 \over T_c^2} \right) ^{\lambda},
\label{anst}
\ee
where $V$ stands for vector mesons, $f_V$ is 
the coupling between the electromagnetic current and the vector
meson field and $\omega_0$ is the continuum threshold. 
Mass of the nucleon also varies with temperature
as Eq.~(\ref{anst}). 
Note that there is no definite reason to believe that all the in-medium
 dynamical quantities are dictated by a single exponent $\lambda$.
 This is the simplest possible ansatz (see~\cite{annals} for a discussion). 
The effective mass of $a_1$
is estimated by using Weinberg's sum rules~\cite{weinberg}.
We have seen earlier that 
the baryon chemical potential has a small effect on the
photon yield~\cite{prc1}. Moreover, in the central
rapidity region the entropy per baryon is 
quite large $\sim 40-50$~\cite{pbm,dumitru}.
Thus the finite baryon density effects 
are neglected here.

We will assume that the produced matter reaches a state of thermodynamic
equilibrium after a proper time $\sim$ 1 fm/c~\cite{bj}. If a
deconfined matter is produced, it evolves in space and time till
freeze-out undergoing a phase transition to hadronic matter in
the process.  The (3+1) dimensional hydrodynamic equations 
have been solved numerically  by the relativistic version of the
flux corrected transport algorithm~\cite{hvg}, assuming boost
invariance in the longitudinal direction~\cite{bj} and cylindrical
symmetry in the transverse plane.
The initial temperature $T_i$
can be related to the multiplicity of the event $dN/dy$ by virtue of the
isentropic expansion as~\cite{hwa},
\be
\frac{dN}{dy}=\frac{45\zeta(3)}{2\pi^4}\pi\,R_A^2 4a_k\,T_i^3\tau_i
\label{dnpidy}
\ee
where $R_A$ is the initial radius
of the system, $\tau_i$ is the initial thermalization time and 
$a_k=({\pi^2}/{90})\,g_k$; $g_k$ being the effective degeneracy 
for the phase $k$ (QGP or hadronic matter). 
The bag model EOS is used for the QGP phase. 
$g_H(T)$, the statistical degeneracy of the hadronic phase,
composed of $\pi$, $\rho$, $\omega$, $\eta$, $a_1$ and
nucleons is a temperature dependent quantity in this case 
and plays an important role in the EOS~\cite{annals}. 
As a consequence the square of sound velocity, 
$c_s^{-2}\,=[(T/g_H)(dg_H/dT)+3] < 1/3$, for the hadronic phase,
indicating non-vanishing interactions among the constituents
(see also~\cite{asakawa}).
The hydrodynamic equations have been solved with
initial energy density, $\epsilon(\tau_i,r)$~\cite{hvg}, obtained from $T_i$ 
through the EOS. We use the following relation
for the initial velocity profile which has been 
successfully used to study transverse momentum spectra of hadrons
~\cite{pbm,uh} and also photons~\cite{kg,dyp},
\be
v_r=v_0\frac{\delta+2}{2}\left(\frac{r}{R_A}\right)^\delta
\label{velprofile} 
\ee
For our numerical calculations we choose $\delta=1$
and sensitivity of the results on $v_0$ will be 
shown. It is observed that the results 
do not change substantially with reasonable 
variation of the parameter $\delta$ for a given value of $v_0$.   
 
\bef
\centerline{\psfig{figure=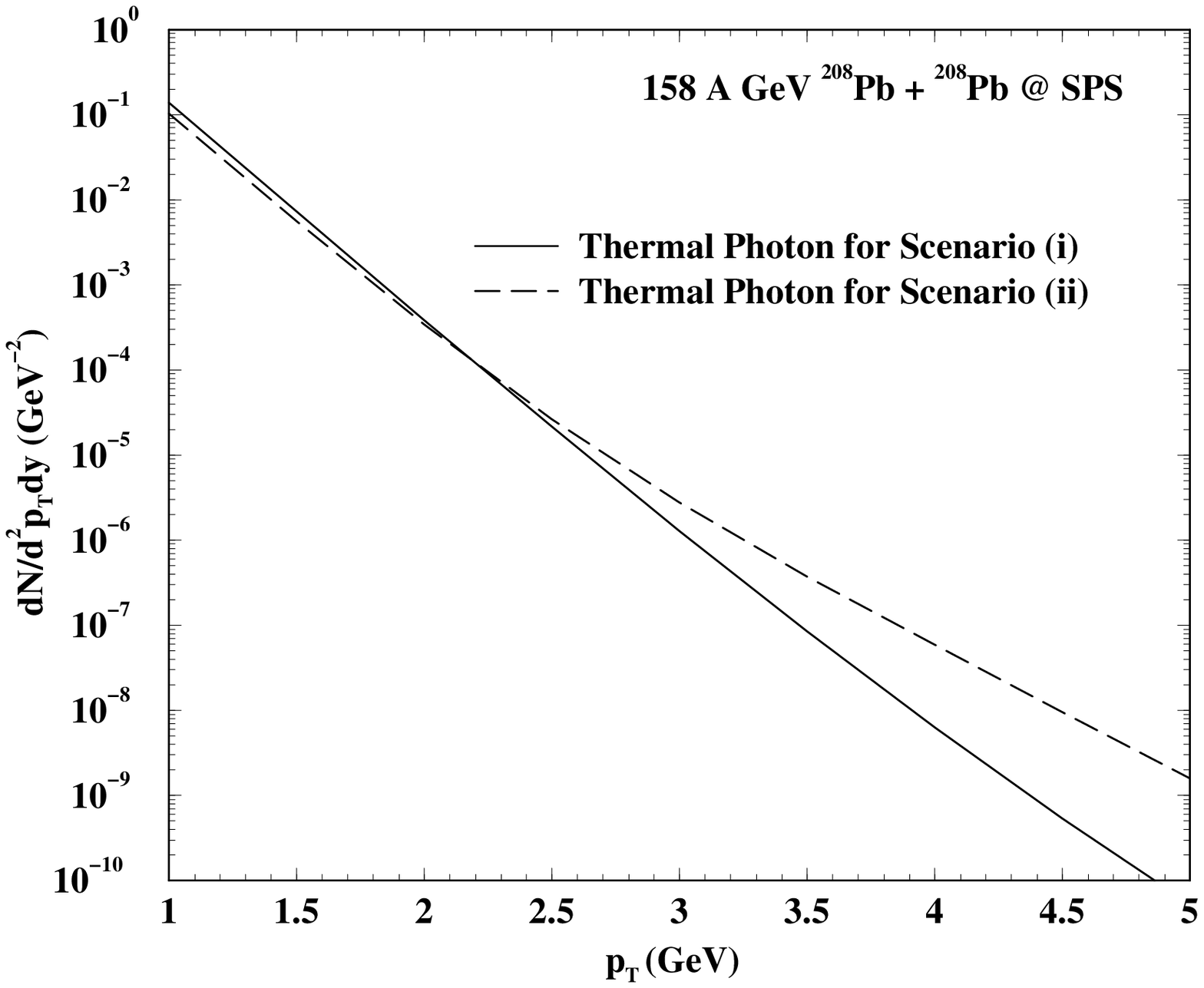,height=6cm,width=8cm}}
\caption{Transverse momentum distribution of thermal photons with
and without medium effects.
}
\label{fig1}
\eef

For central collisions of Pb nuclei at 158 AGeV at 
the CERN-SPS, we assume that QGP is produced at $\tau_i$=1 
fm/c which expands in (3+1) dimension
and undergoes a first order phase
transition to hadronic matter at $T_c$=160 MeV. 
Taking $dN/dy$=700~(last of Ref.~\cite{wa98}), and
$g_k=g_{QGP}$=37 for a two-flavour QGP, the initial temperature $T_i$
comes out as 196 MeV. In a
first order phase transition one has a mixed phase of coexisting QGP
and hadronic matter which persists till the phase transition is over.
Thereafter the hadronic matter expands, cools and freezes out at a
temperature, $T_f$ and radial velocity,
$v_r^f$.  The sum total of the photon yields from
the QGP phase, the mixed phase and the hadronic phase constitutes the
thermal yield. The values of ($T_f, v_r^f$) should in principle be obtained 
from the analysis of hadronic spectra.
However, at present it is not possible to do so without 
ambiguity.  
In Ref.~\cite{epjc2} it is shown that the experimental
data from Pb + Pb collisions allow values of
($T_f, v_r^f$) ranging from (180 MeV,0) to (120 MeV,0.7).
In the present work we take $T_f=120$ MeV and the
initial velocity profile of Eq.(\ref{velprofile}) with 
$v_0$ as a parameter. 

\bef
\centerline{\psfig{figure=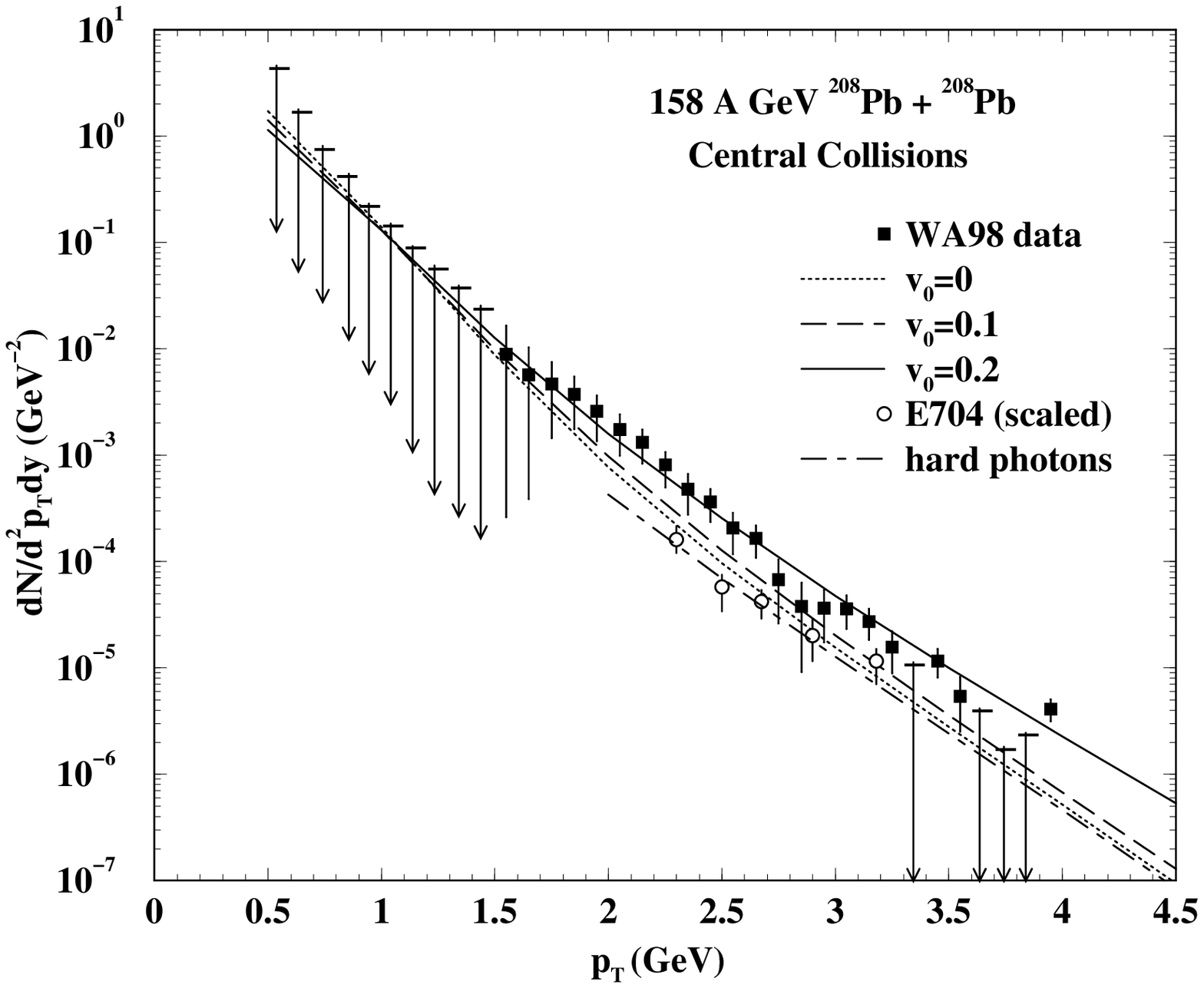,height=6cm,width=8cm}}
\caption{Total photon yield in Pb + Pb collisions
at 158 A GeV at CERN-SPS. The theoretical
calculations contain hard QCD and thermal 
photons. 
The system is formed in the QGP phase with initial temperature 
$T_i=196$ MeV.  
}
\label{fig2}
\eef

In Fig.~\ref{fig1}, 
we show only the {\em thermal photon} spectra originating from 
 quark matter (QM $\equiv$ QGP + QGP part of mixed phase) and
hadronic matter (HM $\equiv$ hadronic part of mixed phase +
hadronic phase) with $v_0=0$. 
The solid and dashed lines correspond to the case 
(i) and (ii) respectively.
  The case with large collisional broadening shows no deviation from  (i). 
 At low $p_T$ the difference between   (i) and (ii) is
negligible because in this region of phase space 
most of the photons are emitted from the late stage of the evolution
where the in-medium effects are small.
The increased photon yield at large $p_T$ is caused by
the enhancement in the Boltzmann factor due to the reduction in meson
(particularly, $\rho$) masses.  
However, in the total photon emission this
 difference of thermal photons at high $p_T$ is masked by the hard
photon contribution. The thermal photon yield with 
hadronic mass variation due to Walecka model or Brown-Rho
scaling~\cite{brpr} ($\lambda=1/6$ in Eq.~(\ref{anst}))
will lie in-between the two curves in Fig.~\ref{fig1}.

In Fig.~\ref{fig2},
 results for the total photon emission is shown for three different values
of the initial transverse velocity with medium effects as in case (ii).
All the three curves represent the sum of the thermal and
the prompt photon contribution which includes possible finite $k_T$
effects of the parton distributions. The later, shown separately by
the dot-dashed line also explains the scaled $pp$ data from E704 
experiment~\cite{e704}. 
We observe that the photon spectra for the initial velocity
profile given by Eq.~(\ref{velprofile}) with  $v_0=0.2$ explains
the WA98 data reasonably well. A similar value of the initial radial
velocity has been obtained in ~\cite{dyp} 
from the analysis of transverse momentum spectra of hadrons and photons
(see also~\cite{kg}). 
It is found that a substantial fraction of the photons come from mixed and 
hadronic phase. The contribution from the QGP phase
is small because of the small life time of the 
QGP ($\sim 1$ fm/c). 
    
\bef
\centerline{\psfig{figure=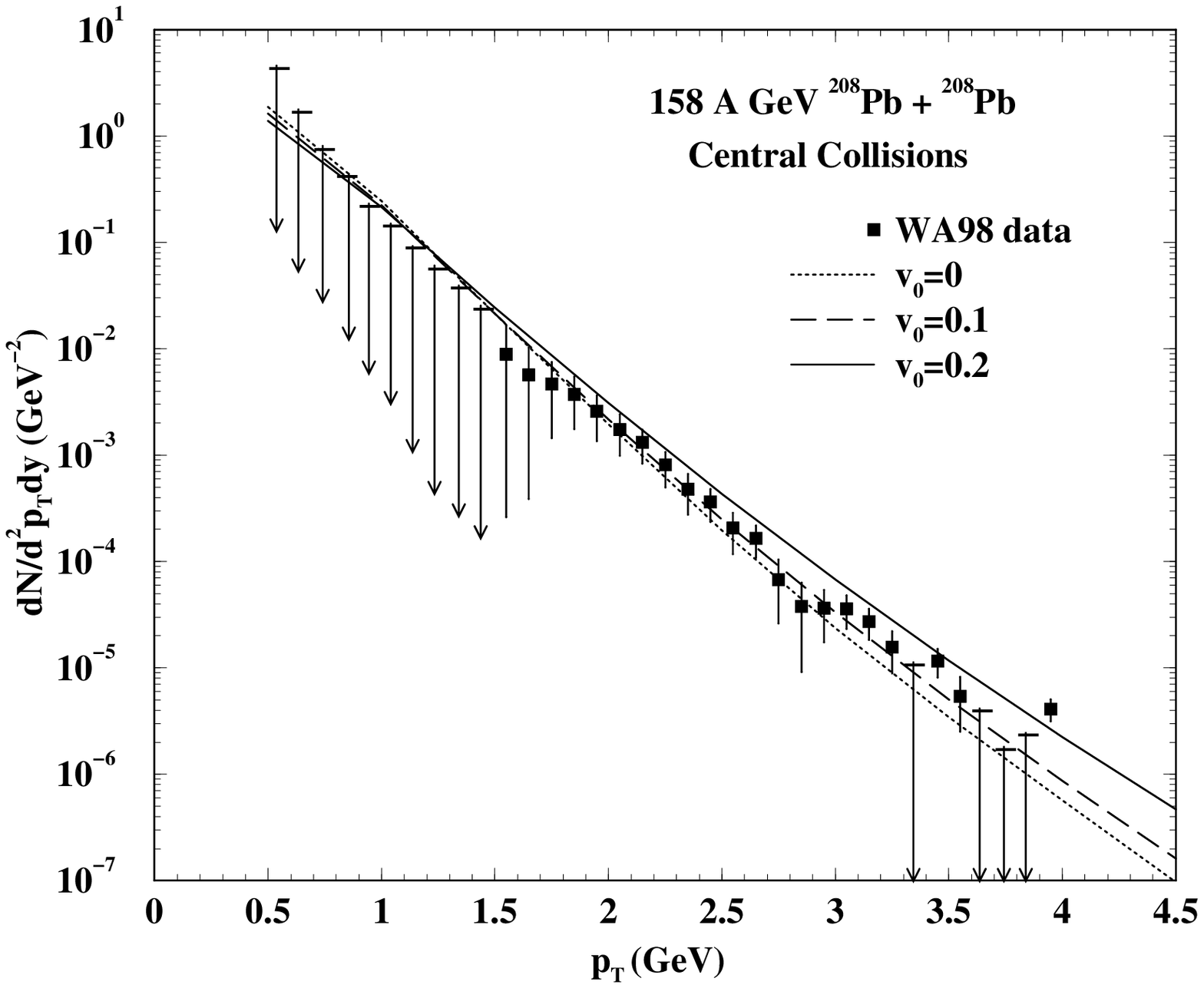,height=6cm,width=8cm}}
\caption{
Total photon yield in Pb + Pb collisions
at 158 A GeV at CERN SPS. The theoretical
calculations contain hard QCD and thermal photons. 
The system is formed in the hadronic phase with 
the hadronic masses approaching zero at initial temperature 
$T_i=205$ MeV.
}
\label{fig3}
\eef

The last statement together with the current uncertainty of the
 critical temperature $T_c$~\cite{lattice} poses the following question: Is
 the  existence of the QGP phase essential to reproduce the 
 WA98 data? To study this problem, we have considered two 
  possibilities: (a) pure hadronic model without medium-modifications,
   and (b) pure  hadronic model  with scaling hypothesis according to
    Eq.(\ref{anst}).
   In the former case,   
$T_i$ is found to be $\sim 250 $ MeV for 
$\tau_i=1$ fm/c and $dN/dy=700$,
 which appears to be too high for the hadrons to survive. Therefore
  this possibility should be excluded.
  On the other hand, 
 the second case with an assumption of $T_i = T_c$ (which is
  just for simplicity) leads to 
  $T_i\sim 205$ MeV, at $\tau_i= 1$ fm/c, which is not unrealistic.
In this case, the hadronic
system expands and cools and ultimately 
freezes out at $T_f$=120 MeV. 
The masses of the vector mesons increase 
with reduction in temperature (due to expansion) according to Eq.(\ref{anst}).
The results of this scenario  
for three values of the initial radial velocity including the prompt photon
contribution are shown in Fig.~\ref{fig3}. 
The experimental data are well reproduced for vanishing 
initial transverse velocity also. 
This indicates that a simple hadronic model is inadequate.
 Either substantial medium modifications
   of hadrons 
  or the formation of QGP in the initial stages is necessary to
   reproduce the data. It is rather difficult to distinguish 
between the two at present.    

We now show our prediction of the photon yield in  central
collisions of Au nuclei at 200 A GeV at RHIC in Fig.~\ref{fig4}
for scenario (ii) with $v_0=0$. 
Enhancement due to intrinsic motion of the partons
has been ignored at RHIC, because at higher beam energies
such effects are small~\cite{cywong}. The initial temperature
and thermalization time are taken as 300 MeV and 0.5 fm/c 
respectively~\cite{dumitru}. 
It is seen that up to $p_T=2$ GeV, most of the photons
are emitted from the thermal source. We also observe
that photons from quark matter make a substantial contribution
to the thermal yield (dash-dotted line in Fig.~\ref{fig4}),
because of the higher initial temperature
and larger lifetime of the QGP phase realized at RHIC compared to
SPS. The effect of the transverse 
expansion on the QGP phase is small. However, photons from
the hadronic phase (short dashed line in Fig.~\ref{fig4}), 
particularly during the late stage of the evolution,
receive a large kick due to the radial expansion and consequently
populate the high $p_T$ region, as shown in Fig.~\ref{fig4}.

\bef
\centerline{\psfig{figure=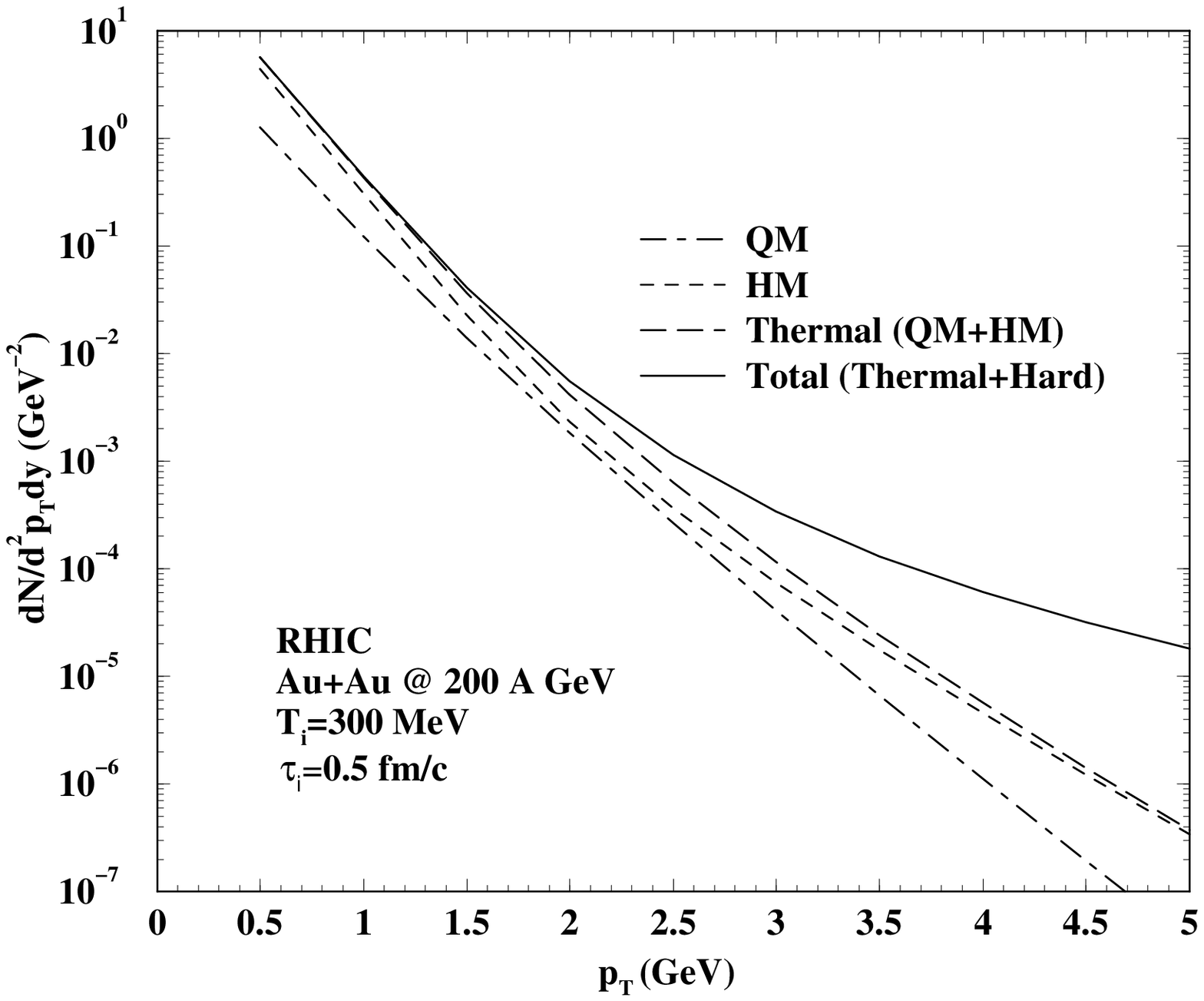,height=6cm,width=8cm}}
\caption{ Photon spectra at RHIC for Au + Au collisions.
}
\label{fig4}
\eef

In summary, we have evaluated the high energy photon yield 
in Pb + Pb collisions at
CERN SPS energies with two different initial conditions.
In the first scenario, we start with the assumption of the formation
of a QGP phase at $T_i\sim 196$ MeV and then the system continues through
mixed phase and hadronic phase before freeze-out. In the second
scenario, we assume a chirally symmetric phase where the hadronic masses
approach zero at a temperature $\sim 205$ MeV and then the system
evolves towards freeze-out.
The effects of the variation of hadronic masses  
on the photon yield have been considered both in the cross section
as well as in the EOS.   
The photon spectra reported by the WA98 collaboration are well
reproduced in both cases.
We thus conclude that the thermal photon 
spectra resulting from the Pb + Pb collisions
at CERN SPS energies are emitted from a source 
with an initial temperature $\sim 200$ MeV. 
A similar value of $T_i$ is also obtained from the
analysis of photon and dilepton data from CERN SPS~\cite{dyp,ctin}.  
At RHIC energies the total yield of photons increases
by an order of magnitude compared to SPS with a
substantial contribution from QM.

In spite of the above encouraging situation, 
a firm conclusion about the formation of the
QGP at SPS necessitates a closer look at some pertinent
but unsettled issues. In the evaluation of the hard photon contribution
the value of the K-factor and the intrinsic $k_T$ distribution
of partons are adjusted so as to reproduce the scaled p-p
data of E704 collaboration.
However, it is extremely important
to know quantitatively the contribution from the hard processes.
The possibility of $A$ and $k_T$ dependence of the K-factor~\cite{jjm},
the energy loss of fast partons, the shadowing of the
structure functions and the broadening of transverse
momentum distribution of partons are some of the vital issues
related to photon production in nucleus-nucleus collisions
which have not been considered in the present work.
Again, the assumption of complete thermodynamic equilibrium
for quarks and gluons may not be entirely
realistic for SPS energies; lack of chemical equilibrium
will further reduce the thermal yield from QGP. 
We have assumed $\tau_i=1$ fm/c at SPS energies, 
which may be considered as the
lower limit of this quantity,  because
the transit time (the time taken by the nuclei to pass
through each other in the CM system) is $\sim$ 1 fm/c at SPS
energies and
the thermal system is assumed to be formed after this time 
has elapsed. 
In the present work, when QGP initial state
is considered, we have assumed a first order phase transition
with bag model EOS for the QGP for its simplicity, although 
it is not in complete agreement with the lattice QCD 
simulations~\cite{lattice}. However, it is difficult to distinguish
among different EOS with the current resolution of the photon data.
As mentioned before, 
there are uncertainties in the value of $T_c$~\cite{lattice},
a value of $T_c\sim 200$ MeV may be considered as an
upper limit.  Moreover, the photon
emission rate from QGP given by Eq.~(\ref{edr}), evaluated in 
Refs.~\cite{kapusta,aurenche}
by resumming the hard thermal loops is strictly valid for $g<<1$ 
whereas the value of $g$ obtained from Eq.~(\ref{strong}) is $\sim 2$
at $T\sim 200 $ MeV. At present it is not clear whether the rate
in Eq.~(\ref{edr}) is valid for such a large vaule of $g$ or not.

\noindent{{\bf Acknowledgement}: 
J.A. is grateful to the Japan
Society for Promotion of Science (JSPS) for financial support.
J.A. and T.H. are also supported by Grant-in-aid for Scientific
Research No. 98360 of JSPS.

\end{document}